\documentstyle[12pt,amsfonts]{article}

\newcommand{\mbf}[1]{\mbox{\boldmath$ #1$}}
\newcommand{\be}{\begin{equation}}
\newcommand{\ee}{\end{equation}}
\newcommand{\ba}{\begin{eqnarray}}
\newcommand{\ea}{\end{eqnarray}}

\begin{document}

\begin{center}
{\LARGE {\bf Lanczos Pseudospectral Propagation Method for Initial-Value
Problems in Electrodynamics of Passive Media}}

\vskip0.5cm {\Large Andrei G. Borisov ${}^{a,}$\footnote{{ email: {\sf %
borisov@lcam.u-psud.fr}}} and Sergei V. Shabanov ${}^{b,}$\footnote{{ %
email: {\sf shabanov@phys.ufl.edu}}}}

\vskip0.4cm ${}^{a}$ {\it Laboratoire des Collisions Atomique et
Mol\'{e}culaires, UMR CNRS - Universit\'{e} Paris-Sud 8625, 91405 Orsay
Cedex, France}

${}^b$ {\it Department of Mathematics, University of Florida, Gainesville,
FL 32611, USA}
\end{center}

\begin{abstract}
Maxwell's equations for electrodynamics of dispersive and absorptive
(passive) media are written in the form of the Schr\"odinger equation with a
non-Hermitian Hamiltonian. The Lanczos time-propagation scheme is modified
to include non-Hermitian Hamiltonians and used, in combination with the
Fourier pseudospectral method, to solve the initial-value problem. The
time-domain algorithm developed is shown to be unconditionally stable.
Variable time steps and/or variable computational costs per time step with
error control are possible. The algorithm is applied to study transmission
and reflection properties of ionic crystal gratings with cylindric geometry
in the infra-red range.
\end{abstract}

\newpage

{\bf 1. Introduction}. There is a demand for fast time-domain solvers of the
Maxwell's equations to model dynamics of broad band electromagnetic pulses
in dispersive and absorptive media. This is driven by various applications
that include photonic devices, communications, and radar technology, amongst
many others. Efficiency, accuracy, and stability are the key criteria of
choosing a concrete algorithm for specific applications. In the past
decades, pseudospectral methods of solving the initial value problem for
differential equations have been under intensive study \cite{BoydBook}.
Because of their high efficiency and accuracy, they have replaced finite
differencing approaches in many traditional applications as well as
scientific simulations, e.g., in quantum chemistry \cite{chem}.
Unconditionally stable pseudospectral algorithms are particularly attractive
for numerical simulations.

In the present paper we develop an unconditionally stable time-domain
algorithm for solving the initial value problem for Maxwell's equations in
dispersive and absorptive (passive) media with sharp interfaces
(discontinuities of medium parameters). It is a time-stepping algorithm that
is based on the Hamiltonian formalism for electrodynamics of passive
continuous media, the Lanczos propagation scheme \cite{lanczos,park}, and
the Fourier pseudospectral method \cite{kosloff}. Apart from the
unconditional stability, the algorithm has a dynamical control of accuracy,
which allows one to automatically optimize computational costs with error
control at each time step.

We apply the algorithm to the scattering of broad band electromagnetic
pulses on gratings, the photonic devices that currently attract lots of
attention because of their transmission and reflection properties \cite{gr}.
As for the passive medium, we choose an ionic crystal material. From the
numerical point of view, the model of the dielectric permeability of such a
material is rather representative and used in the vast number of
applications. From the physical point of view, the interest to gratings and
photonic crystals made of this kind of material is due two types of effects
in interaction with electromagnetic radiation: The structural and
polaritonic ones \cite{Ionic1,Ionic2}. We show that in the infrared range
the reflection and transmission properties of ionic crystal gratings change
significantly in narrow frequency ranges due to structural and polaritonic
resonances. Structural resonances are associated with the existence of
trapped (quasistationary) electromagnetic modes supported by the grating
geometry (guided wave resonances) \cite{DielGratings}. Polaritonic
resonances are associated with dispersive properties of the material. Such
resonances appear when the incident radiation can cause polaritonic
excitations in the medium. From the macroscopic point of view, this occurs
in the anomalous dispersion region of the dielectric constant.

{\bf 2. Basic equations}. Maxwell's equations in passive media can be
written in the form of the Schr\"{o}dinger equation in which the wave
function is a multidimensional column, composed of electromagnetic field
components and the medium polarization, and the Hamiltonian is, in general,
non-Hermitian when attenuation is present. The initial-value problem (the
time evolution of an electromagnetic pulse) is then solved by finding the
fundamental solution (the evolution operator kernel) for the Schr\"{o}dinger
equation. Here this idea is applied to the ionic crystal material whose
dielectric properties at the frequency $\omega $ are described by the
dielectric constant 
\begin{equation}  \label{eps}
\varepsilon (\omega )=\varepsilon _{\infty }+\frac{(\varepsilon
_{0}-\varepsilon _{\infty })\omega _{T}^{2}}{\omega _{T}^{2}-\omega
^{2}-i\eta \omega }\ ,  \label{1}
\end{equation}
where $\varepsilon _{\infty ,0}$ are constants, $\omega _{T}$ is the
resonant frequency, and $\eta $ is the attenuation. In particular,
transmission and reflection properties of the periodic grating structure of
circular parallel cylinders made of such a material are studied.

Let ${\bf P}$ be a dispersive part of the total polarization vector of the
medium. Then ${\bf D}=\varepsilon _{\infty }{\bf E}+{\bf P}$, where ${\bf D}$
and ${\bf E}$ are the electric induction and field, respectively. By using
the Fourier transform, it is straightforward to deduce that ${\bf P}$
satisfies the second-order differential equation 
\begin{equation}
\ddot{{\bf P}}+\eta \dot{{\bf P}}+\omega _{T}^{2}{\bf P}=\varepsilon
_{\infty }\omega _{p}^{2}{\bf E}\ ,  \label{2}
\end{equation}
where the overdot denotes the time derivative, $\omega _{p}^{2}=(\varepsilon
_{0}-\varepsilon _{\infty })\omega _{T}^{2}/\varepsilon _{\infty }$ if $%
\varepsilon _{0}-\varepsilon _{\infty }$ is positive, otherwise, $\omega
_{p}^{2}\rightarrow -\omega _{p}^{2}$ in (\ref{2}). Equation (\ref{2}) must
be solved with the zero initial conditions, ${\bf P}=\dot{{\bf P}}=0$ at $%
t=0 $.

Define a set of auxiliary fields ${\bf Q}_{1,2}$ by ${\bf P}=\sqrt{%
\varepsilon _{\infty }}\omega _{p}{\bf Q}_{1}/\omega _{T}$ and $\dot{{\bf Q}}%
_{1}=\omega _{T}{\bf Q}_{2}$. For non-magnetic media ($\mu =1$), the
Maxwell's equations and (\ref{2}) can be written as the Schr\"{o}dinger
equation, $i\dot{\psi}(t)=H\psi (t)$, in which the wave function and the
Hamiltonian are defined by 
\begin{equation}
\psi =\pmatrix{ \varepsilon_\infty^{1/2}{\bf E}\cr {\bf B}\cr {\bf Q}_1\cr
{\bf Q}_2}\ ,\ \ \ H=\pmatrix{0& ic
\varepsilon_\infty^{-1/2}\mbf{\nabla}\times &0& -i\omega_p\cr
-ic\mbf{\nabla}\times \varepsilon_\infty^{-1/2}&0&0&0\cr 0&0&0&i\omega_T\cr
i\omega_p&0& -i\omega_T&-i\eta }\ ,  \label{3}
\end{equation}
where $c$ is the speed of light in the vacuum, and ${\bf B}$ is the magnetic
field. A solution to the initial-value problem is given by $\psi (t)=\exp
(-itH)\psi (0)$. Boundary conditions at medium interfaces are enforced
dynamically, that is, parameters of the Hamiltonian $H$ are allowed to be
discontinuous functions of position. In particular, $\varepsilon _{\infty ,0}
$ are set to one in the vacuum, and to some specific values in the medium in
question (see Section 4).

The norm of the wave function, $\Vert \psi \Vert ^{2}=\int d{\bf r}\psi
^{\dagger }\psi $, is proportional to the total electromagnetic energy of
the wave packet \cite{bs,Split}. When the attenuation is not present, $\eta
=0$, the Hamiltonian is Hermitian, $H^{\dagger }=H$, relative to the
conventional scalar product in the space of square integrable functions, and
the norm (energy) is conserved.

{\bf 3. The algorithm}. The Lanczos propagation scheme is based on a
polynomial approximation of the short-time fundamental solution for the Schr%
\"{o}dinger equation, $\psi (t+\Delta t)=\exp (-i\Delta tH)\psi (t)$ \cite
{park,review}. If the exponential is expanded into the Taylor series and the
latter is truncated at the order $O(\Delta t^{n})$, then the approximation
of $\psi (t+\Delta t)$ belongs to the Krylov space ${\sf K}_{n}={\rm Span}%
\,\{\psi (t),H\psi (t),...,H^{n-1}\psi (t)\}$. Let $P_{n}$ be a projection
operator of the original Hilbert space onto the Krylov space, $%
P_{n}^{2}=P_{n}$ and $P_{n}^{\dagger }=P_{n}$. Let $P_{n}\psi =\psi _{(n)}$.
The exact solution of the initial-value problem is approximated by a
solution of the corresponding initial value problem in ${\sf K}_{n}$, $\psi
_{(n)}(t+\Delta t)=\exp (-i\Delta tH^{(n)})\psi _{(n)}(t)$, where $%
H^{(n)}=P_{n}HP_{n}$. The accuracy of the approximation is of order $%
O(\Delta t^{n})$. Hence, for practical needs, it is sufficient to choose $n$
not so large (typically, $n\leq 9$). Then $H^{(n)}$ is a small matrix whose
exponential is computed by direct diagonalization. If the Hamiltonian is
Hermitian, then the corresponding matrix is symmetric tridiagonal in the
Lanczos basis for ${\sf K}_{n}$ \cite{lanczos,park}. Therefore the
propagation scheme is unitary, that is, the energy (norm) of the initial
wave packet is preserved, which also implies the unconditional stability of
the algorithm. Another important feature of the Lanczos propagation scheme
for Hermitian Hamiltonians is the dynamical control of accuracy \cite
{park,review}.

For non-Hermitian Hamiltonians, one possibility is to use the split method
in combination with the Lanczos propagation scheme for the Hermitian part of
the Hamiltonian \cite{Lanczos1}. While the unconditional stability is
maintained, the accuracy is limited by the accuracy of the split
approximation of the infinitesimal fundamental solution, typically, by $%
O(\Delta t^{3})$. An alternative, mathematically sound, and more accurate
procedure is based on the use of the dually orthonormal Lanczos basis in
which $H^{(n)}$ retains the tridiagonal complex symmetric structure \cite
{lanczos}. In the particular case of a complex symmetric $H$, the use of the
dual Krylov space can be avoided. There is an orthogonal Lanczos basis for $%
{\sf K}_{n}$ in which $H^{(n)}$ is also complex symmetric tridiagonal, but
the orthogonality is now understood with respect to a new scalar product
(without complex conjugation of vectors) \cite{vorst}. However, in both the
cases, the projector $P_{n}$ is no longer Hermitian. This has an unpleasant
consequence. One can show that, while the accuracy remains of the order $%
O(\Delta t^{n})$, the unconditional stability is typically lost. The
algorithm is only conditionally stable. A more detailed study of such
schemes will be given elsewhere. Here we focus on developing an
unconditionally stable algorithm. For this purpose we construct an
orthonormal basis for ${\sf K}_{n}$ itself, ignoring the dual Krylov space
and making $P_{n}$ Hermitian. In this basis $H^{(n)}$ has a Hessenberg form,
that is, it is upper-triangular with one extra non-zero lower superdiagonal 
\cite{vorst}. A direct diagonalization of such a matrix is still not
expensive for small $n$. The dynamical control of accuracy is also preserved.

Let $\psi _{j}$, $j=0,1,...,n-1$, form an orthonormal basis for ${\sf K}_{n}$
that is constructed as follows. Set $\phi _{0}=\psi (t)$ and $\psi _{0}=\phi
_{0}/\Vert \phi _{0}\Vert $. Compute $h_{00}^{(n)}=(\psi _{0},H\psi _{0})$
where $(\cdot \,,\,\cdot )$ denotes the conventional scalar product in the
Hilbert space of square integrable functions. For $j=1,2,...,n-1$, compute 
\begin{eqnarray}
\phi _{j} &=&H\psi _{j-1}-\sum_{k=0}^{j-1}h_{kj-1}^{(n)}\psi _{k}\ ,
\label{5} \\
\psi _{j} &=&\phi _{j}/\Vert \phi _{j}\Vert \ ,\ \ \ \ \
h_{jj-1}^{(n)}=(\psi _{j},H\psi _{j-1})\ ,
\end{eqnarray}
and, for $k=0,1,...,j$, compute 
\begin{equation}  \label{hn}
h_{kj}^{(n)}=(\psi _{k},H\psi _{j})\ .  \label{6}
\end{equation}
By construction, $(\psi _{j},\psi _{k})=\delta _{jk}$. Let $\psi
_{(n)}(t+\Delta t)=P_{n}\psi (t+\Delta t)=\sum_{k=0}^{n-1}c_{k}(\Delta
t)\psi _{k}$ where $c_{k}(\Delta t)=(\psi _{k},\psi (t+\Delta t))$. Note
that the basis functions $\psi _{j}$ depend on $t$ and so do $c_{j}$. For
brevity of notations, the dependence on $t$ of all quantities involving $%
\psi _{j}$ is not explicitly shown in what follows. The expansion
coefficients $c_{k}$ are united into a complex column ${\bf c}$ with $n$
entries. It is straightforward to infer that ${\bf c}(\Delta t)$ satisfies
the Schr\"{o}dinger equation $i\dot{{\bf c}}=h^{(n)}{\bf c}$ with the
initial condition $c_{k}(0)=\delta _{k0}$. The matrix $h^{(n)}$ has a
Hessenberg form and is the projected Hamiltonian $H^{(n)}$ in the basis
constructed. Thus ${\bf c}(\Delta t)=\exp (-i\Delta th^{(n)}){\bf c}(0)$.
The exponential of the Hessenberg matrix $h^{(n)}$ is computed by direct
diagonalization.

To compute the basis functions, multiple actions of the Hamiltonian $H$ on
the current wave function $\psi (t)$ are required. This is done by the
Fourier pseudospectral method on a finite grid \cite{kosloff}. To suppress
reflections of the wave packet from grid boundaries, absorbing boundary
conditions are enforced by a conducting layer placed at the grid edges (see,
e.g., \cite{CAPS}). The position dependence of the conductivity $\sigma $ is
adjusted to suppress reflections with desired accuracy in the frequency
range of interest. The Hamiltonian $H$ is modified accordingly. In the upper
left corner of $H$ in (\ref{3}), the function $-4\pi i\sigma $ is inserted
instead of zero.

Let us discuss the stability of the algorithm. In a time-stepping algorithm,
an amplification matrix $G(\Delta t)$ is defined by $\psi(t+\Delta
t)=G(\Delta t)\psi(t)$. The algorithm is unconditionally stable if $%
\|G^N(\Delta t)\|\leq const$ uniformly for all integers $N>0$, $\Delta t
\geq 0$ and all other parameters characterizing the system \cite{richt}. The
norm of an operator is defined by $\|G\|=\sup \|G\psi\|/\|\psi\|$. For any $%
H $ the following decomposition holds: $H=H_0 +iV$ where $H_0$ and $V$ are
Hermitian. Clearly, $V$ is responsible for attenuation in any physically
reasonable model of a passive medium. The norm of the initial-value problem
solution decreases with time if $V$ is negative semidefinite, that is, for
any $\psi$, $(\psi, V\psi)\leq 0$. Now we prove that the Lanczos algorithm
with the Hessenberg projected Hamiltonian (\ref{hn}) is unconditionally
stable, provided $V$ of the total Hamiltonian is negative semidefinite.

Observe that $\|\exp(-i\Delta t H)\| \leq 1$ for $\Delta t >0$ and all
parameters of $H$ for which $V$ remains negative semidefinite. The
amplification matrix has the form $G(\Delta t) = \exp(-i\Delta t H^{(n)})$.
Thanks to the Hermiticity of $P_n$, it is sufficient to show that $V^{(n)}
=P_nVP_n$ is negative semidefinite because the latter implies that $%
\|G^N(\Delta t)\| \leq \|G(\Delta t)\|^N\leq 1$ uniformly for all integers $%
N>0$, $\Delta t >0$, and all parameters of $H$. For any $\psi$, the
following chain of equalities holds, $(\psi, V^{(n)}\psi) = (\psi,
P_nVP_n\psi) = (\psi_{(n)}, V\psi_{(n)}) \leq 0$. In the first equality, the
definition of $V^{(n)}$ has been used, in the second one, the Hermiticity of
the projection operator has been invoked, and the final inequality is valid
since $V$ is negative semidefinite. The proof is completed.

The most expensive operation of the algorithm is the action of the
Hamiltonian on the wave function. Bearing also in mind the accumulation of
round-off errors, when computing powers of $H$ with broad spectrum (cf. \cite
{lanczos}), this implies that the dimension of the Krylov space has to be as
small as possible at each time step while controlling the approximation
error. In the case of Hermitian Hamiltonians, the $n$ needed at a specified
time step can be deduced from the condition \cite{park} that $%
|c_{n-1}(\Delta t)|^{2}\leq \epsilon $ where $\epsilon $ is a small number.
This condition is based on the fact that $c_{n-1}(\Delta t)\sim O(\Delta
t^{n-1})$. Note that $c_{n-1}$ determines the weight of $H^{n-1}\psi (t)$ in
the Taylor expansion of $\psi (t+\Delta t)$. In our algorithm for a
non-Hermitian $H$, the time evolution of the vector ${\bf c}$ is generated
by a Hessenberg matrix $h^{(n)}$. By examining the Taylor expansion of the
exponential in $\exp (-i\Delta th^{(n)}){\bf c}(0)$ it is easy to convince
oneself that $c_{n-1}(\Delta t)\sim O(\Delta t^{n-1})$ remains valid thanks
to $c_{j}(0)=\delta _{j0}$. Thus, the same dynamical accuracy control can be
used. In particular, in our simulations the dimension of the Krylov space at
each time step is determined by $|c_{n-3}|^{2}+|c_{n-2}|^{2}+|c_{n-1}|^{2}%
\leq \epsilon $ at a fixed $\Delta t$ to control weights of three highest
Krylov vectors in the approximate solution, and $\epsilon \sim 10^{-14}$
with $\Delta t$ being fixed. Note, however, that both the propagation
parameters $n$ and $\Delta t$ can be varied at each time step to minimize
computational costs at the given accuracy level.

{\bf 4. Ionic crystal gratings}. The algorithm has been applied to simulate
the scattering of broad band electromagnetic (laser) pulses on a grating
structure consisting of circular parallel ionic crystal cylinders
periodically arranged in vacuum. Our primary interest is to study the effect
of trapped modes (guided wave resonances) and polaritonic excitations on
transmission and reflection properties of the grating in the infrared range.
The dielectric function of the ionic crystal material is approximated by the
single oscillator model (\ref{eps}). Following the work \cite{Ionic1}, we
choose the parameters representative for the beryllium oxide: $\varepsilon
_{\infty }=2.99$, $\varepsilon _{0}=6.6$, $\omega _{T}=87.0$ meV, and the
damping $\eta =11.51$ meV. The packing density $R/D_{g}=0.1$, where $R$ is
the radius of cylinders and $D_{g}$ is the grating period, has been kept
fixed in simulations. The cylinders are set parallel to the $y$ axis. The
structure is periodic along the $x$ axis, and the $z$ direction is
transverse to the grating. A Gaussian wave packet propagating along the $z$
axis is used as an initial configuration. It is linearly polarized with the
electric field oriented along the $y$ axis, i.e., parallel to the cylinders
(the so called TM polarization). The frequency resolved transmission and
reflection coefficients are obtained via the time-to-frequency Fourier
transform of the signal on ``virtual detectors'' placed at some distance in
front and behind the periodic structure \cite{VirtualDet}. The zero
diffraction mode is studied here for wavelengths $\lambda \geq D_{g}$
so that reflected and transmitted beams propagate along the $z$-axis.
Similar to our previous works \cite{bs,Lanczos1} we use a change of
variables in both $x$ $(x=f_{1}(\xi ))$ and $z$ $(z=f_{2}(\zeta ))$
coordinates to enhance the sampling efficiency in the vicinity of medium
interfaces. A typical size of the mesh corresponds to $-17D_{g}\leq z\leq
15D_{g}$, and $-0.5D_{g}\leq x\leq 0.5D_{g}$ with, respectively, $384$ and $%
64$ knots. Note that, because of the variable change, a uniform mesh in the
auxiliary coordinates $(\xi ,\zeta )$ corresponds to a non-uniform mesh in
the physical $(x,z)$ space.

Two types of resonances are expected for the gratings studied here.
Structure resonances are associated with the existence of guided wave modes 
\cite{DielGratings,Lanczos1}. They are characteristic for periodic
dielectric gratings and, in the absence of losses, lead to the 100\%
reflection within a narrow frequency interval(s) for wavelengths $\lambda
\sim D_{g}$. The second type of resonances arise because of polariton
excitations for wavelengths $\lambda \sim D_{T}=2\pi c/\omega _{T}$.
Calculations have been done for different values of $D_{g}$ so that the
polaritonic excitation can be tuned through out the wavelength range of
interest ($\lambda /D_{g}\geq 1$) by changing the ratio $D_{T}/D_{g}$.

In Fig. 1 we show the results obtained for the transmission (blue curves)
and reflection (red curves) coefficients for the beryllium oxide gratings
characterized by the period $D_{g}$ such that $D_{T}/D_{g}=0.5$, $2.5$, and $%
4$ as indicated in the figure. The results are presented as a function of
the radiation wavelength measured in units of the grating period. Note that
the logarithmic scale is used for the horizontal axis in order to improve
the resolution at small wavelengths. Consider first the following two
limiting cases. According to (\ref{eps}), for short wavelengths $\lambda \ll
D_{T}$ ($D_{g}\ll D_{T}$), the medium behaves as as a dielectric with $%
\varepsilon \approx \varepsilon _{\infty }$. In the long wavelength limit $%
\lambda \gg D_{T}$ ($D_{g}\gg D_{T}$), the medium responds as a dielectric
material characterized by $\varepsilon \approx \varepsilon _{0}$. In Fig. 1
the dashed and solid black curves represent the reflection coefficient of
the grating made of a lossless, non-dispersive dielectric with $\varepsilon
=\varepsilon _{\infty }$ and $\varepsilon =\varepsilon _{0}$, respectively.
In agreement with the previously published results \cite{Lanczos1}, the
reflection coefficient in these two cases reaches $1$ within a narrow
frequency range for $\lambda \sim D_{g}$. This resonant pattern is
associated with the so-called Wood anomalies \cite{Wood}, and can be
explained by the existence of trapped modes or guided wave resonances \cite
{DielGratings,bs}. The width of the resonances is determined by the lifetime
of the corresponding quasi-stationary trapped mode which is a standing wave
along the $x$ axis and is excited by the incoming wave. The width increases
with $\varepsilon $ while the resonant wavelength gets redshifted, which
explains the difference between the dashed and solid black curves ($%
\varepsilon _{0}>\varepsilon _{\infty }$).

Now we turn to the discussion of the effects due to dispersive properties of
the ionic crystal material. For $D_{T}/D_{g}=0.5$, the resonant excitation
of polaritons is impossible within the range of wavelengths of interest, and
the dielectric constant is close to $\varepsilon _{0}$. The result for the
reflection coefficient in this case is similar to the data shown by the
black solid curve. However, there is an essential difference as compared to
the case of a lossless, non-dispersive dielectric grating. Indeed, for a
lossless medium the sum of the reflection and transmission coefficients must
be one, which is not the case for the beryllium oxide model because of the
damping (the dashed-dotted green curve). The maximal loss of energy
corresponds to the resonant wavelength. It is easily understood because the
trapped mode remains in contact with the material much longer than the main
pulse, and, therefore, can dissipate more energy.

For $D_{T}/D_{g}=2.5$, two resonances emerge leading to the enhanced
reflection within the corresponding frequency ranges. The one at $\lambda
/D_{g}\sim 2.5$, i.e., $\lambda \sim D_{T}$, is associated with polaritonic
excitations of the ionic crystal. The resonance at $\lambda \sim D_{g}$ is a
structure resonance. As follows from (\ref{eps}), the dielectric constant in
this case approaches $\varepsilon _{\infty }$ for small wavelengths $\lambda
\sim D_{g}$. Then the width and position of the structure resonance are
close to the data given by the dashed black curve. The imaginary part of the
dielectric function is large enough through out the entire wavelength range
to produce a substantial energy loss at both the resonances.

Finally, for $D_{T}/D_{g}=4$ the polariton excitation appears at $\lambda
\sim 4D_{g}$ and the two resonances are well separated. The structure
resonance at $\lambda \sim D_{g}$ closely matches the result for a lossless,
non-dispersive dielectric grating characterized by $\varepsilon =\varepsilon
_{\infty }$. Observe that the reflection coefficient is close to 1 in this
case and the energy loss is small because the imaginary part of $\varepsilon
(\omega )$ is small far from $\omega =\omega _{T}$.

Figs 2 and 3 show the reflection and transmission coefficients of the
grating as functions of the incident radiation wavelength and the grating
period $D_{g}$. The polariton resonance wavelength $D_{T}=2\pi c/\omega _{T}$
and the packing density $R/D_{g}$ are kept fixed. The results for the two
limiting cases in which $\varepsilon =\varepsilon _{0}$ and $\varepsilon
=\varepsilon _{\infty }$ are represented by the left most and right most
colored columns, respectively. The resonance pattern of the system is
clearly visible, in particular, the transformation of the structure
resonance at $\varepsilon =\varepsilon _{0}$ into the polaritonic one. Thus,
by increasing the ratio $D_{T}/D_{g}$, the ``broad'' structure resonance
associated with $\varepsilon =\varepsilon _{0}$ is turned into the
polaritonic resonance and follows the diagonal of the plot ($\lambda
/D_{g}=D_{T}/D_{g}$). At the same time, starting approximately with $%
D_{T}/D_{g}=2$, the ``narrow'' structure resonance associated with $%
\varepsilon =\varepsilon _{\infty }$ emerges and fully develops for $%
D_{T}/D_{g}=4$.

Finally we would like to show the sensitivity of the results to the
attenuation in the present system. In Fig. 4, the transmission and
reflection coefficients are presented for two different choices of the
attenuation $\eta $ in (\ref{eps}). The geometry of the grating structure is
set by $D_{T}/D_{g}=2.5$. The upper panel of the figure corresponds to $\eta
=11.51$ meV as used throughout this paper. The lower panel of the figure
corresponds to the damping reduced by the factor of $20:$ $\eta \rightarrow
\eta /20$. Overall features are qualitatively the same in both the cases.
Thus, both the structure resonance at $\lambda \sim D_{g}$ and polaritonic
resonance at $\lambda \sim 2.5D_{g}$ are present, accompanied by the reduced
transmission and enhanced reflection. As the wavelength increases from the
structure resonance, the reflectivity of the grating drops to zero. Its
subsequent onset for $\lambda >1.683D_{g}$ is linked with the metal-type
behavior of the ionic crystal ($\varepsilon $ becomes negative). The
characteristic frequency for the ``metallization'' can be deduced from the
Lyddane-Sachs-Teller relation $\omega _{L}=\omega _{T}\sqrt{\varepsilon
_{0}/\varepsilon _{\infty }}$, leading to $\lambda _{L}=1.683D_{g}$ for $%
\lambda _{T}=2.5D_{g}$. Despite these common features, the reduction of the
attenuation leads to essential changes. In contrast to the upper panel of
the figure, for $\eta \rightarrow \eta /20$ the transmission coefficient
reaches nearly $0$ at both the resonances, and the reflection coefficient is
close to $1$. Moreover, new structures appear in the polaritonic resonance
for $\lambda =D_{T}$, i.e., as $\varepsilon $ changes from large negative to
large positive values. These structures are completely washed out for the
medium with large damping. This result indicates the importance of accurate
modelling of losses in polaritonic media in order to make reliable
predictions of transmission and reflection properties of grating structures
and photonic crystals.

{\bf 5. Conclusions}. We have developed an unconditionally stable
(time-domain) algorithm for initial value problems in electrodynamics of
inhomogeneous, dispersive, and absorptive media. The method is based on the
three essential ingredients: ({\it i}) the Hamiltonian formalism in
electrodynamics of passive media, ({\it ii}) the Lanczos propagation scheme,
modified to account for attenuation, and ({\it iii}) the Fourier
pseudospectral method on non-uniform grids induced by change of variables to
enhance the sampling efficiency in the vicinity of sharp inhomogeneities of
the medium. Apart from the unconditional stability, the algorithm allows for
a dynamical accuracy control, meaning that the two propagation parameters,
the dimension of the Krylov space and the time step, may automatically be
adjusted to minimize computational costs in due course of simulations, while
controlling error.

The algorithm has been tested by simulating the scattering of infrared
electromagnetic pulses on periodic gratings composed of parallel cylinders
that are made of the ionic crystal material. The Lorentz model describing
dielectric properties of such a material is rather representative and used
to model a vast variety of dielectric materials. Our results demonstrate the
role of structure (or guided wave) resonances and polaritonic excitations
for the transmission and reflection properties of grating structures. The
results are also shown to be sensitive to the attenuation of polaritonic
media.

{\bf Acknowledgments}. S.V.S. thanks the LCAM of the Univestity of Paris-Sud
and, in particular, Dr. V. Sidis for support and warm hospitality extended
to him during his stay in Orsay. S.V.S. is also grateful to Dr. R. Albanese
(US Air Force Brooks Research Center, TX), Profs. J.R. Klauder and T. Olson
(University of Florida) for the continued support of this project. The
authors thank Dr. D. Wack (KLA-Tencor, San Jose, CA) for stimulating and
supporting the work on this project.

\newpage {\bf Figure captions}

Fig. 1.\qquad Calculated zero-order reflection (red curves) and transmission
(blue curves) coefficients for the ionic crystal grating described in the
text. The results are presented as a function of the incident radiation
wavelength measured in units of the grating period $D_{g}$. Different panels
of the figure correspond to different values of the grating period as
compared to the resonance wavelength for the polaritonic excitation of the
material, $D_{T}=2\pi c/\omega _{T}$. The dashed and solid black curves
represent the reflection coefficient calculated for the grating made of a
lossless, non-dispersive dielectric characterized by $\varepsilon
=\varepsilon _{\infty }$ and $\varepsilon =\varepsilon _{0}$, respectively.
The sum of the reflection and transmission coefficients is shown as the
dashed-dotted green curve. Its deviation from $1$ represents the
electromagnetic energy loss because of the attenuation. \bigskip 

Fig. 2.\qquad The zero-order transmission coefficient for the ionic crystal
grating described in the text as a function of the incident radiation
wavelength and the grating period. The horizontal axis represents the ratio $%
D_{T}/D_{g}$ of the resonance wavelength for the polaritonic excitation of
the material $D_{T}=2\pi c/\omega _{T}$ and the grating period $D_{g}$. The
vertical axis represents the incident radiation wavelength $\lambda $
measured in units of $D_{g}$. Color codes used for the plot are shown in the
inset.\bigskip 

Fig. 3.\qquad The zero-order reflection coefficient for the ionic crystal
grating described in the text as a function of the incident radiation
wavelength and the grating period. The horizontal axis represents the ratio $%
D_{T}/D_{g}$ of the resonance wavelength for the polaritonic excitation of
the material $D_{T}=2\pi c/\omega _{T}$ and the grating period $D_{g}$. The
vertical axis represents the incident radiation wavelength $\lambda $
measured in units of $D_{g}$. Color codes used for the plot are shown in the
inset.\bigskip 

Fig. 4.\qquad Calculated zero-order reflection (red curves) and transmission
(dashed blue curves) coefficients for the ionic crystal grating. The sum of
the reflection and transmission coefficients is shown as the dashed-dotted
green curve. The geometry of the grating structure is set by $D_{T}/D_{g}=2.5
$. The upper panel of the figure corresponds to the attenuation $\eta =11.51$
meV as used throughout the paper. The lower panel of the figure corresponds
to the damping reduced by the factor of $20:$ $\eta \rightarrow \eta /20$.
The vertical black line defines the resonant wavelength $\lambda =D_{T}=2\pi
c/\omega _{T}$.

\end{document}